\documentclass[aps,prd,amssymb,nofootinbib,twocolumn,epsf,floatfix,
               superscriptaddress]{revtex4}
\usepackage[usenames]{color}
\usepackage[normalem]{ulem} 
\usepackage{graphicx} 
\usepackage{subfigure} 
\usepackage{amssymb}  
\usepackage{amsmath}
\usepackage{mathrsfs}
\usepackage{epstopdf}

\makeatletter
\renewcommand{\p@subfigure}{\thefigure}
\makeatother

\begin{document}
 

\title{Uncertainty Quantification for the Relativistic Inverse Stellar
Structure Problem}

\author{Lee Lindblom}
\affiliation{Department of Physics, University of California at San
  Diego, San Diego, California, 92093 USA}

\author{Tianji Zhou}
\affiliation{Department of Physics and Astronomy, Haverford College,
             Haverford, Pennsylvania, 19041 USA}

\date{\today}
 
\begin{abstract}
  The relativistic inverse stellar structure problem determines the
  equation of state of the stellar matter given a knowledge of
  suitable macroscopic observable properties (e.g. their masses and
  radii) of the stars composed of that material.  This study
  determines how accurately this equation of state can be determined
  using noisy mass and radius observations.  The relationship between
  the size of the observational errors and the accuracy of the
  inferred equation of state is evaluated, and the optimal number of
  adjustable equation of state parameters needed to achieve the
  highest accuracy is determined.
\end{abstract} 
 
\maketitle

\section{Introduction}
\label{s:Introduction}

The quantity and quality of astrophysical observations of the masses
and radii of neutron stars has improved significantly in the past
decade~\cite{Ozel2016a, Riley2019, Miller2019, Miller2021, Biswas2021,
  Vinciguerra2024, Salmi2024, Choudhury2024, Chatziioannou2024}.
Masses have now been measured at the few percent level for dozens of
neutron stars, and both mass and radius have been measured for a few
neutron stars at the 10-20\% level.  It is well known that a knowledge
of the masses and radii of neutron stars can be used to infer the
equation of state of the high density material in the cores of these
stars~\cite{Lindblom1992}.  The purpose of this paper is to apply a
simple uncertainty quantification analysis~\cite{Smith2014} to the
relativistic inverse stellar structure problem, i.e. the problem of
determining the high density neutron-star equation of state from a
knowledge of the masses and radii of those stars.  The goal of this
analysis is to determine the relationship between the accuracy of the
available mass-radius data with the accuracy of the equation of state
that can be inferred from those data.

Section~\ref{s:InverseProblem} of this paper describes the method of
solving the relativistic inverse stellar structure problem used in
this study.  This method constructs a parametric representation of the
high-density equation of state by fixing its parameters to minimize
the differences between the resulting model neutron stars and the
observed mass-radius data~\cite{Lindblom2012, Lindblom2014,
  Lindblom2014a}.  The particular parametric equation of state
representation used in this study is a causal spectral representation
with basis functions constructed from Chebyshev
polynomials~\cite{Lindblom2024b}.  The basic properties of this
equation of state representation are summarized in
Appendix~\ref{s:CausalRepresentations}.

Section~\ref{s:NoisyMRData} constructs the collections of mock noisy
mass-radius data used in this study to evaluate the accuracy of the
equations of state determined from them.  These mock noisy data are
constructed here by adding random errors of various sizes to the exact
masses and radii computed from the GM1L nuclear-theory based
neutron-star equation of state~\cite{spinella2017:PHD,
  Glendenning1991, Typel2010}.  Four collections of mock data, each
containing 1000 noisy mass-radius curves, are constructed with random
fractional error amplitudes 0.1\%, 1\%, 10\%, and 20\%.

Section~\ref{s:Uncertainty Quantification} solves the relativistic
inverse stellar structure problem using the mock noisy mass-radius
data prepared in Sec.~\ref{s:NoisyMRData}.  Parameterized equations of
state using different numbers of spectral parameters,
$N_\mathrm{parms}=1,...,5$ are determined for each noisy mass-radius
curve by solving the inverse stellar structure problem as described in
Sec.~\ref{s:InverseProblem}.  The accuracy of these parametric
equations of state are evaluated by measuring the differences between
them and the exact GM1L equation of state from which the mock data are
constructed.  The dependence of these equation of state errors on the
accuracy of the mock observational data is then evaluated for equation
of state representations with different numbers of spectral
parameters.

Section~\ref{s:Discussion} presents a brief summary of the results of
the uncertainty quantification analysis presented in this study, along
with a discussion of some interesting implications of these results.

\section{Inverse Stellar Structure Problem}
\label{s:InverseProblem}

The matter in the cores of neutron stars is driven to very high
temperatures by the gravitational collapse of the matter that forms
these stars.  These temperatures briefly become high enough to disrupt
the atomic nuclei of this matter, so this material is expected to have
universal thermodynamic properties determined by nuclear physics and
not on the prior thermodynamic history of material from which it
formed.  All neutron stars are therefore expected to be composed of
material having the same high-density equation of state.  The
relativistic inverse stellar structure problem determines this
equation of state from a knowledge of the macroscopic observable
properties of neutron stars, e.g. their masses and radii or tidal
deformabilities~\cite{Lindblom2014a}.  This section summarizes the
particular solution to this problem used in this
study~\cite{Lindblom2012, Lindblom2014}.

The relativistic inverse stellar structure problem is solved here by
representing the unknown equation of state parametrically,
$\epsilon=\epsilon(p,\upsilon_a)$, where $\epsilon$ is the total
energy density of the material, $p$ is the pressure, and the
$\upsilon_a$ for $a=1,...,N_\mathrm{parms}$ are adjustable parameters.
These parameters are fixed by making the stellar models based on this
equation of state match the observational mass-radius data
$\{M_i,R_i\}$ for $i=1,...,N_\mathrm{MR}$ as closely as possible.

The particular equation of state representation used in this study is
based on a Chebyshev polynomial spectral
expansion~\cite{Lindblom2024b} which is described in
Appendix~\ref{s:CausalRepresentations}.  The Oppenheimer-Volkoff
relativistic stellar structure equations~\cite{Oppenheimer1939} are
solved using the equation of state with specified values of
$\upsilon_a$ to determine the masses, $M(p_c^i,\upsilon_a)$, and
radii, $R(p_c^i,\upsilon_a)$, of the stellar models with central
pressures $p_c^i$.  Given a collection of mass-radius observables,
$\{M_i,R_i\}$ for $i=1,...,N_\mathrm{MR}$, the values of the equation
of state parameters $\upsilon_a$ are adjusted to make the stellar
model properties $\{M(p_c^i,\upsilon_a), R(p_c^i,\upsilon_a)\}$ match
the observables $\{M_i,R_i\}$ as closely as possible.  This is done by
minimizing the error function $\chi^2$, defined by,
\begin{eqnarray}
  \chi^2(p_c^i,\upsilon_a) &=& \frac{1}{N_\mathrm{MR}}
  \sum_{j=1}^{N_\mathrm{MR}}\left\{\left[\log\left(
    \frac{M(p_c^j,\upsilon_a)}{M_j}\right)\right]^2\right.
  \nonumber\\
  &&\left.\qquad\qquad+\left[\log\left(
    \frac{R(p_c^j,\upsilon_a)}{R_j}\right)\right]^2\right\},
    \label{e:chi_MR}
\end{eqnarray}
with respect to the equation of state parameters, $\upsilon_a$, as
well as the central pressures, $p_c^i$, of the stars with
observational data points, $\{M_i,R_i\}$.  The equation of state
determined by the parameters $\upsilon_a$ that minimize $\chi^2$
is therefore an approximate solution to the inverse stellar structure
problem.

The mock observational data points $\{M_i,R_i\}$ used in this study
were constructed from a particular nuclear-theory based equation of
state.  Since the exact equation of state is known in this case, it is
possible to evaluate the accuracy of the parametric equation of state,
$\epsilon=\epsilon(p,\upsilon_a)$, determined by the solution to the
inverse stellar structure problem.  This is done by evaluating the
equation of state error function, $\Delta$, defined by,
\begin{equation}
  \Delta^2(\upsilon_a)=\frac{1}{N_{EOS}}\sum_{k=1}^{N_{EOS}}\left[
    \log\left(\frac{\epsilon(p_k,\upsilon_a)}{\epsilon_k}\right)\right]^2,
  \label{e:Delta_EOS}
\end{equation}
where $\{\epsilon_k,p_k\}$ are points from the exact equation of state
table used to generate the mock observational data points,
$\{M_i,R_i\}$.  The Chebyshev polynomial based equation of state
representations used in this study have been shown to provide
convergent representations for a wide range of neutron-star equations
of state~\cite{Lindblom2024b}.~\footnote{The mock mass-radius data
analyzed in this study are from the contiguous branch of stable
neutron-stars. The Chebyshev based representations have been tested
for a wide range of neutron-star equations of state, including those
with a range of phase transitions that may exist in these stars.} This
study explores the extent to which approximate solutions to the
inverse stellar structure problem are also convergent in the sense
that they produce equation of state error functions
$\Delta(\upsilon_a)$ that decrease toward zero as the number of
parameters $\upsilon_a$ increases.

\section{Noisy Mass-Radius Data}
\label{s:NoisyMRData}

Mock noisy observational data $\{M_i,R_i\}$ are constructed for this
study by adding random errors to a collection of points $\{\tilde M_i,
\tilde R_i\}$ from the exact mass-radius curve associated with a known
equation of state.  These random errors are parameterized by an error
amplitude $\mathcal{A}$ and a random phase variable $\delta$.  The
noisy mock data $\{M_i,R_i\}$ are generated from $\{\tilde M_i,\tilde
R_i\}$:
\begin{eqnarray}
  M_i &=& (1+\delta\mathcal{A})\,\tilde M_i,
  \label{e:Mtilde}\\
  R_i &=& (1+\delta\mathcal{A})\,\tilde R_i.
  \label{e:Rtilde}
\end{eqnarray}
The random phase variables $\delta$ are uniformly distributed over the
domain $-1\leq \delta \leq 1$.   They are computed for this study using
the random number generator ran2 from Ref.~\cite{numrec_f}.  The error
amplitude $\mathcal{A}$ is the maximum fractional error of each point
in each noisy mock observational mass-radius curve, $\{M_i,R_i\}$.

\begin{figure}[!h]
    \includegraphics[width=0.4\textwidth]
                    {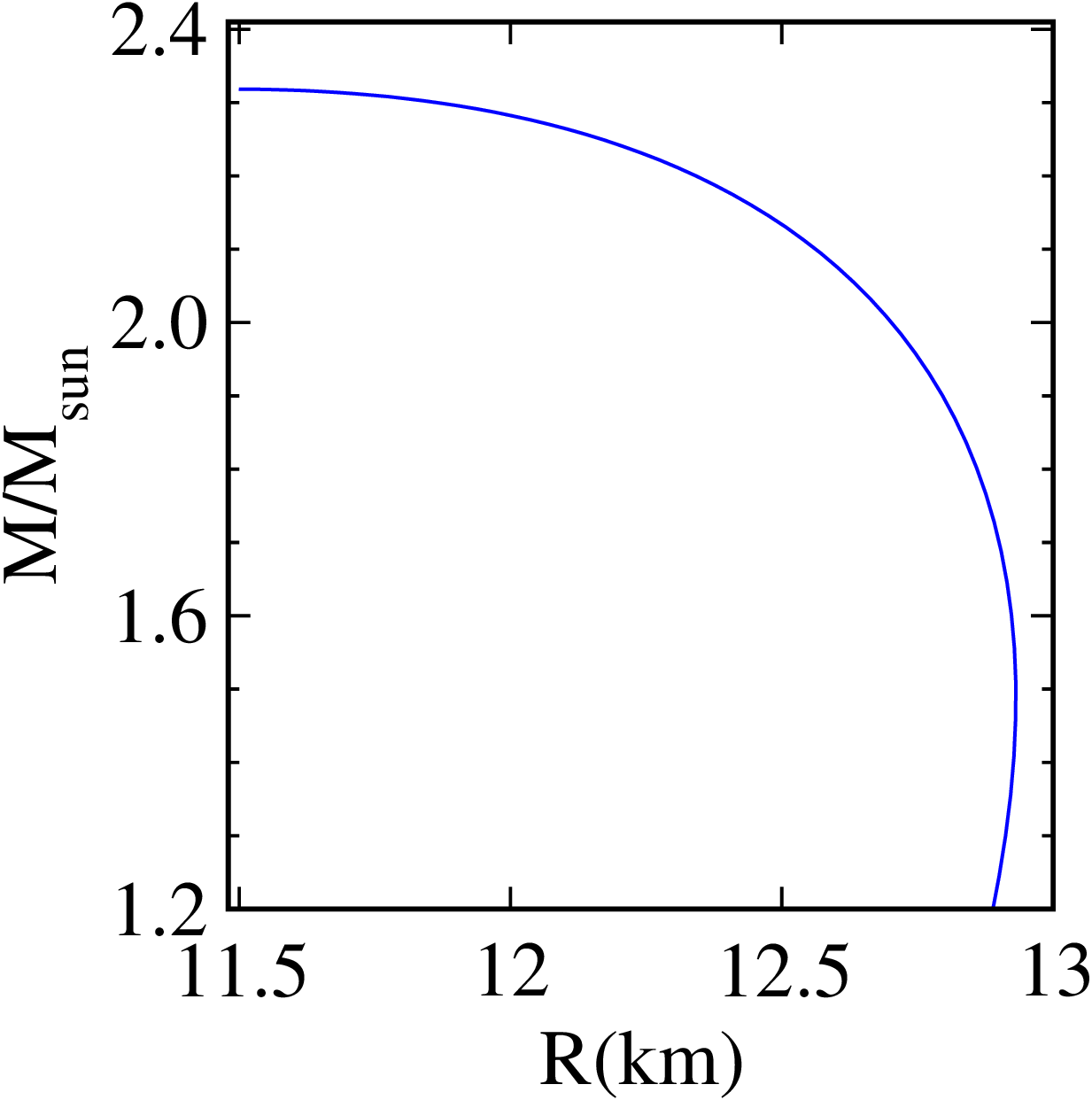}
          \caption{Neutron-star model masses and radii computed from
            the GM1L nuclear-theory based equation of state.
          \label{f:MR_GM1L}}
\end{figure}

The reference equation of state used to compute the mock $\{M_i,R_i\}$
data for this study is the nuclear-theory based equation of state
GM1L.  This equation of state was constructed in
Ref.~\cite{spinella2017:PHD} from the relativistic mean field GM1
equation of state of Ref.~\cite{Glendenning1991} by adjusting the
slope of the symmetry energy to agree with the established value,
$L=55$ MeV, using the formalism developed in Ref.~\cite{Typel2010}.
The masses and radii of the neutron-star models computed from this
equation of state are illustrated in Fig.~\ref{f:MR_GM1L} for the
models with masses in the astrophysically interesting range $M\geq 1.2
M_\mathrm{sun}$.

\begin{figure}[!h] 
    \includegraphics[width=0.4\textwidth]
                    {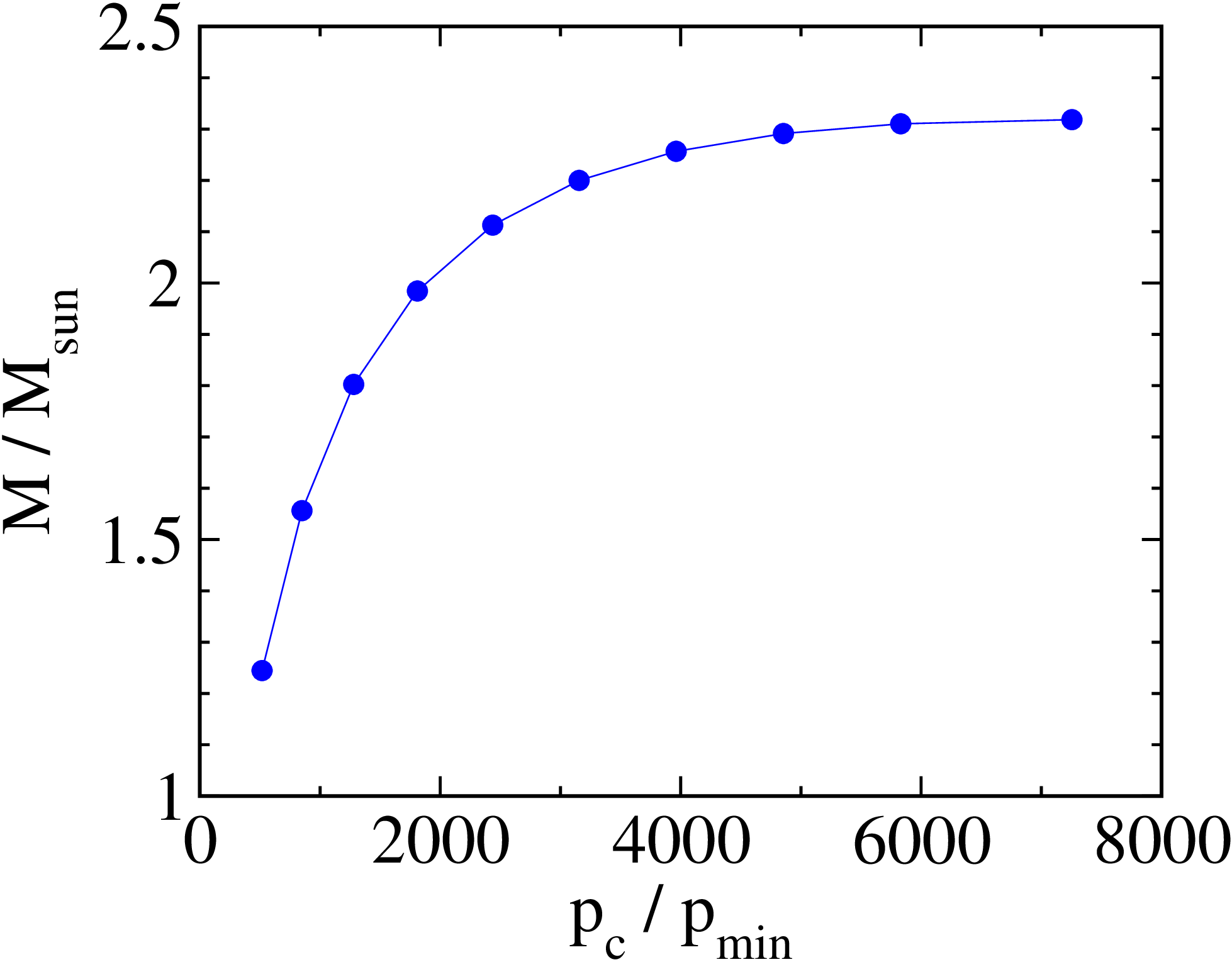}
          \caption{Neutron-star masses as functions of their central
            pressures for the ten $\{\tilde M_i,\tilde R_i\}$
            reference data points used in this study.  The constant
            $p_\mathrm{min}=1.20788\times10^{32}$ erg/cm${}^3$ used to
            scale the pressures in this figure is the lower limit of
            the domain on which spectral representations are used for
            the high density equation of state.
          \label{f:MP_DATA}}
\end{figure}

The reference $\{\tilde M_i,\tilde R_i\}$ data points used in this
study were selected from the mass-radius curve illustrated in
Fig.~\ref{f:MR_GM1L}.  The number of data points $N_\mathrm{MR}$ was
chosen to be $N_\mathrm{MR}=10$ for this study.  This aspirational
choice is somewhat larger than the currently available number of high
quality data points, but is within the range of what might become
available in the not too distant future.  The distribution of these
$N_\mathrm{MR}=10$ data points are also chosen aspirationally for this
study.  Rather than choosing these points randomly from the
astrophysically relevant range of masses, $1.2 M_\mathrm{sun} \leq M
\leq M_\mathrm{max}$, they are chosen here to sample models whose
central pressures are relatively uniformly distributed along the high
density equation of state curve.  Figure~\ref{f:MP_DATA} illustrates
the $\tilde M_i$ chosen here as a function of the central pressures of
those stellar models.  The equation of state errors achieved with
these aspirational choices for the mock observational data are likely
to be somewhat smaller than could be achieved with a smaller number of
randomly distributed data points.  The numerical values of these exact
$\{\tilde M_i,\tilde R_i\}$ data points are listed in
Table~\ref{t:MR_Data_Exact}.
\begin{table}[!h]
\begin{center}
  \caption{$\{\tilde M_i,\tilde R_i\}$ data for the neutron-star
    models computed from the GM1L nuclear-theory equation of state.
    These reference $\{\tilde M_i,\tilde R_i\}$ values are used to
    generate the noisy data sets for the uncertainty quantification
    analysis in this study.
    \label{t:MR_Data_Exact}}
  \begin{ruledtabular}
  \begin{tabular}{c c | c c}
    \qquad$\tilde M/M_\mathrm{sun}$ \qquad &  \qquad$\tilde R$(km)\qquad\qquad
     &   \qquad$\tilde M/M_\mathrm{sun}$ \qquad &  \qquad$\tilde R$(km)\qquad
    \vspace{0.07cm} \\
    \hline
1.244629 & 12.90062 & 2.200004 & 12.34184 \\
1.556269 & 12.92859 & 2.256765 & 12.13824 \\
1.802454 & 12.85875 & 2.291466 & 11.93727 \\
1.984403 & 12.71947 & 2.310322 & 11.74336 \\
2.112802 & 12.54009 & 2.318418 & 11.49951 \\
  \end{tabular}
  \end{ruledtabular}
\end{center}
\end{table}

\begin{figure}[!t]
    \includegraphics[width=0.4\textwidth]
                    {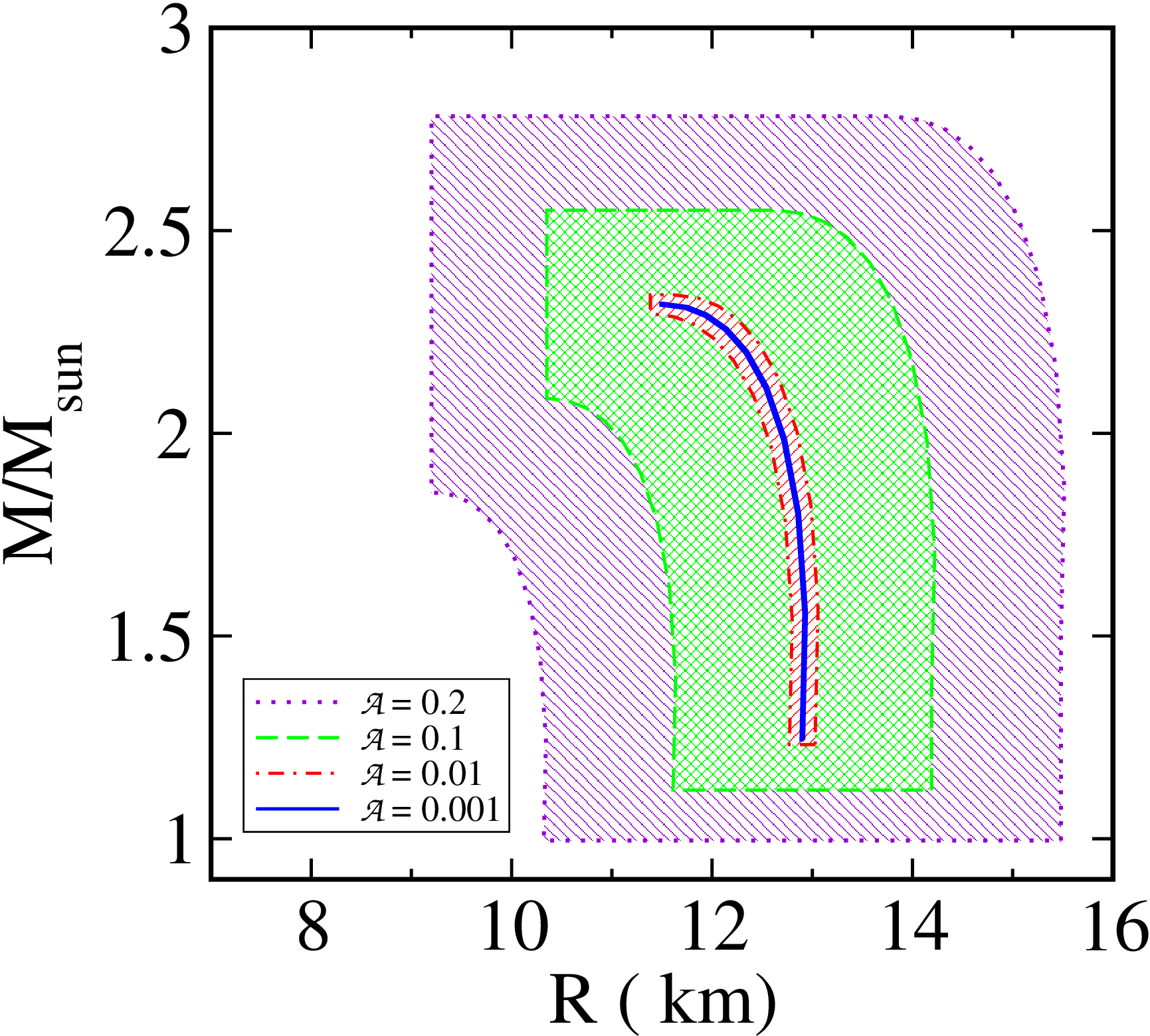}
          \caption{Regions of the mass-radius plane occupied by the
            perturbed $\{M_i,R_i\}$ data sets used in this study.
            Random perturbations with amplitudes
            $\mathcal{A}=\{0.2,0.1,0.01,0.001\}$ were added to the
            exact $\{\tilde M_i,\tilde R_i\}$ points from the GM1L
            equation of state to construct data sets with 1000
            perturbed $\{M_i,R_i\}$ curves for each of the four error
            amplitudes.
          \label{f:MR_DATA_RANGES}}
\end{figure}

Starting from the exact reference $\{\tilde M_i,\tilde R_i\}$ given in
Table~\ref{t:MR_Data_Exact}, collections of noisy mock data
$\{M_i,R_i\}$ were constructed using Eqs.~(\ref{e:Mtilde}) and
(\ref{e:Rtilde}) for four different fractional error amplitudes:
$\mathcal{A}=\{0.2,0.1,0.01,0.001\}$.  The larger values,
$\mathcal{A}=\{0.2,0.1\}$ , are more or less at the currently
achievable observational error levels, while the smaller values
$\mathcal{A}=\{0.01,0.001\}$ were included to explore how accurately
the equation of state might be determined if/when more accurate data
become available.  Mock observational data collections, each
containing 1000 $\{M_i,R_i\}$ noisy mass-radius curves, were
constructed for each of these error amplitudes:
$\mathcal{A}=\{0.2,0.1,0.01,0.001\}$.  Figure~\ref{f:MR_DATA_RANGES}
illustrates the regions of the mass-radius plane occupied by the
points in these four mock observational data collections.
%

\section{Uncertainty Quantification}
\label{s:Uncertainty Quantification}

This section describes the uncertainty quantification analysis of the
relativistic inverse stellar structure problem performed for this
study.  This analysis begins by solving the inverse stellar structure
problem outlined in Sec.~\ref{s:InverseProblem} for each collection of
mock observational data $\{M_i,R_i\}$ described in
Sec.~\ref{s:NoisyMRData}.  These solutions provide equations of state
for each data set in the collections defined by their error
amplitudes, $\mathcal{A}=\{0.2,0.1,0.01,0.001\}$.  Evaluating the
relationship between the accuracies of these equations of state and
the accuracies of the mock observational data used to compute them
determines the uncertainty quantification for this problem.

The first step in this analysis is to minimize the differences between
the model observables $\{M(p_c^i,\upsilon_a), R(p_c^i,\upsilon_a)\}$
and the mock observational data $\{M_i,R_i\}$ as measured by the
mass-radius fitting function $\chi$ defined in Eq.~(\ref{e:chi_MR}). These
minimizations have been carried out for this study for each of the
mock observational data sets using equation of state representations
with $N_\mathrm{parms}=1,...,5$ spectral parameters.  The numerical
calculations used in this study were performed using two independent
codes to confirm the accuracy of the results. The $\chi^2$
minimizations were carried out numerically using a Fortran
implementation of the Levenburg-Marquardt algorithm as described in
Ref.~\cite{numrec_f}, and using the scipy.optimize.least\_squares
implementation in Python~\cite{2020SciPy-NMeth}.  The resulting
minimal values of $\chi$ obtained by these independent codes agree to
within a few percent.

\begin{figure}[!t]
    \includegraphics[width=0.4\textwidth]
                    {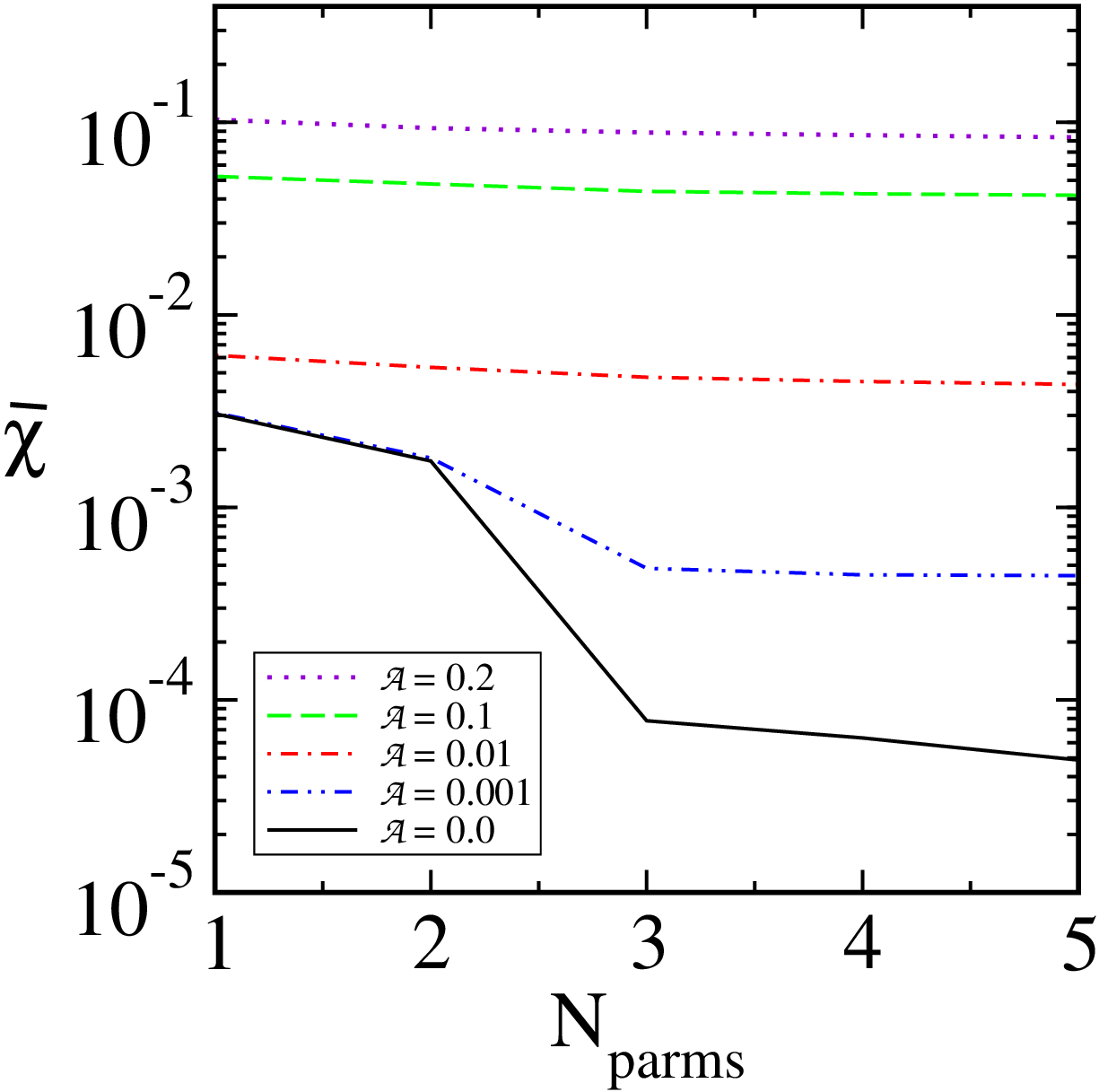}
          \caption{Values of $\bar\chi$, the mass-radius fitting errors
            defined in Eq.~(\ref{e:chi_MR}) averaged over the
            collection of mock observational $\{M_i,R_i\}$ data,
            as functions of $N_\mathrm{parms}$, the number of spectral
            parameters used to model the high density equation of
            state.  
          \label{f:chi_Average}}
\end{figure}

Figure~\ref{f:chi_Average} illustrates how well the model observables
are able to fit the collections of mock noisy data with error
amplitudes $\mathcal{A}=\{0.2,0.1,0.01,0.001\}$.  The quantity
$\bar\chi$ is the average of the minimum values of $\chi$ over the the
1000 $\{M_i,R_i\}$ data sets with error amplitude $\mathcal{A}$.
Figure~\ref{f:chi_Average} illustrates $\bar\chi$ for each collection
of mock data as functions of the number of spectral parameters used to
compute the model observables.  These results show that these minimum
$\bar\chi$ values are about half the values of the observational
errors for the $\mathcal{A}=\{0.2,0.1,0.01\}$ data collections, as
well as the $N_\mathrm{parms}=3,..,5$ cases for the
$\mathcal{A}=0.001$ collection.  The solid (black) curve in
Fig.~\ref{f:chi_Average} is the graph of the minimum $\chi$ values for
the exact reference $\{\tilde M_i,\tilde R_i\}$ data.  These exact
results, labeled $\mathcal{A}=0.0$ in Fig.~\ref{f:chi_Average}, show
that the minimum values of $\bar \chi$ for the $N_\mathrm{parms}=1,2$
cases of the $\mathcal{A}=0.001$ collection are limited by the
accuracy of the spectral representation of the equation of state
rather than the accuracy of the observational $\{M_i,R_i\}$ data.

\begin{figure}[!t]
    \includegraphics[width=0.4\textwidth]
                    {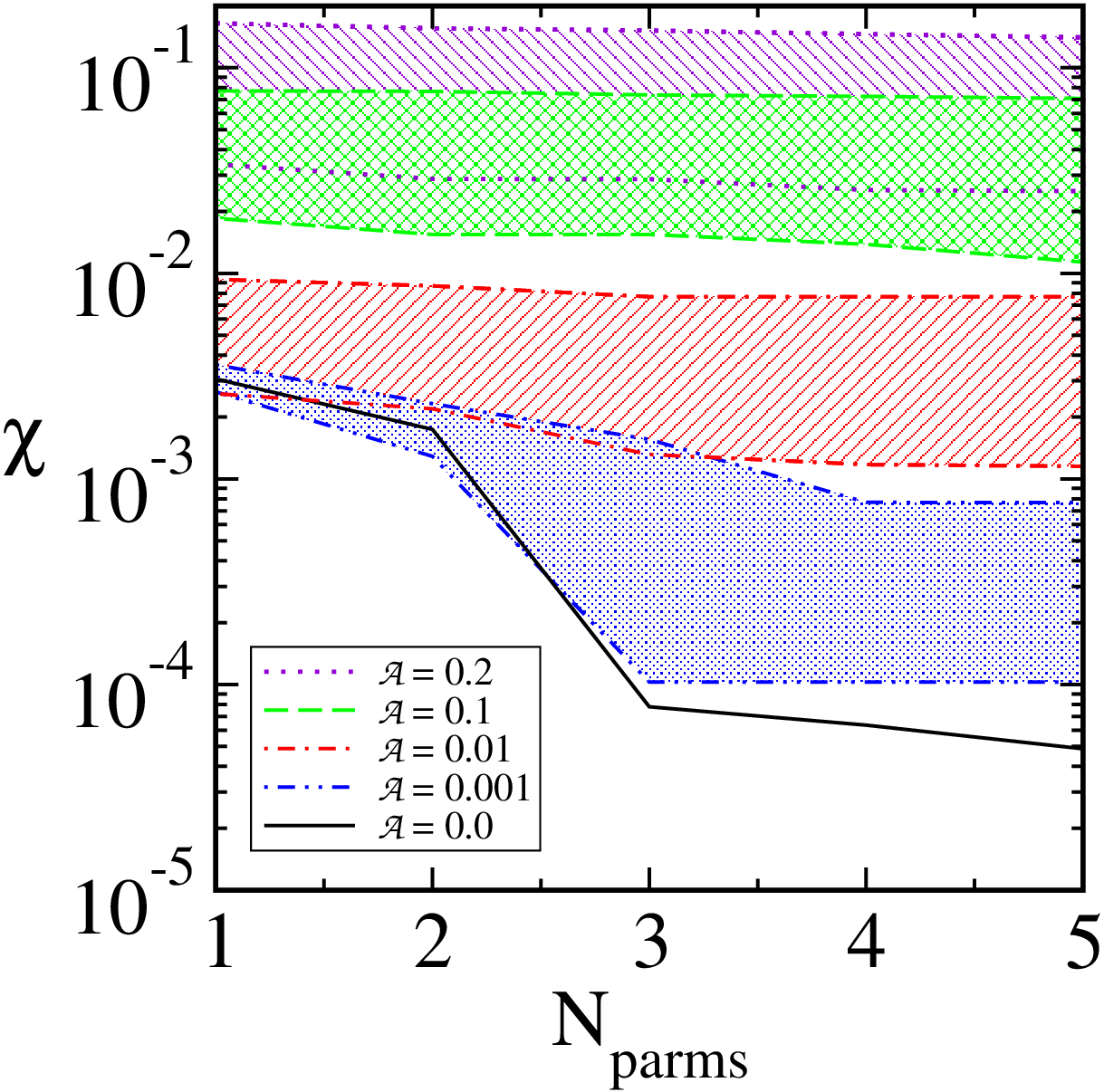}
          \caption{Ranges of the minimum values of $\chi$, the mass-radius
            fitting errors defined in Eq.~(\ref{e:chi_MR}), for each
            collection of mock observational $\{M_i,R_i\}$ data as
            functions of $N_\mathrm{parms}$, the number of spectral
            parameters used to model the high-density equation of
            state.
          \label{f:chi_Ranges}}
\end{figure}

Figure~\ref{f:chi_Ranges} illustrates the range of $\chi$ values
obtained for each collection of noisy $\{M_i,R_i\}$ data.  Note that
the upper limits of these $\chi$ ranges are smaller than the
observational data errors for the $\mathcal{A}=\{0.2,0.1,0.01\}$ data
collections, and for the $N_\mathrm{parms}=4,5$ cases of the
$\mathcal{A}=0.001$ collection.  These results show that the
neutron-star models constructed here do a good job of representing
noisy observational $\{M_i,R_i\}$ data at an accuracy level
commensurate with the amplitude of the observational errors.

\begin{figure}[!h]
    \includegraphics[width=0.4\textwidth]
                    {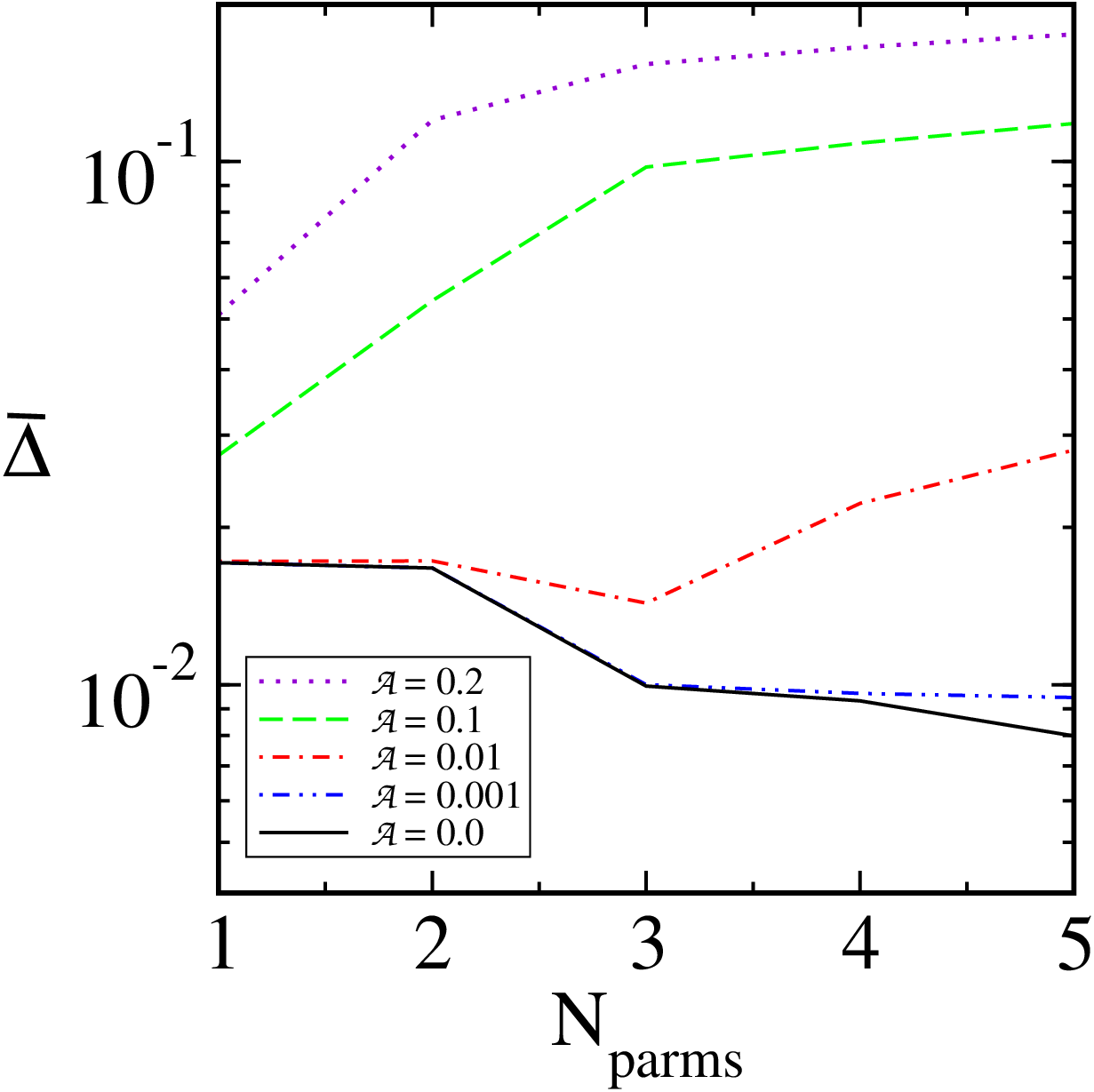}
          \caption{Equation of state errors $\bar\Delta$ averaged over
            each collection of 1000 $\{M_i,R_i\}$ data sets are shown
            as functions of the number of spectral parameters
            $N_\mathrm{parms}$. The fitting error $\Delta$ is also
            shown for the equation of state obtained using the exact,
            $\mathcal{A}=0.0$, $\{\tilde M_i,\tilde R_i\}$ data set as
            the solid (black) curve.
          \label{e:Delta_Average}}
\end{figure}

The next step is to evaluate the accuracy of the equations of state
determined by the spectral parameters, $\upsilon_a$, that minimize the
mass-radius fitting errors $\chi$.  These accuracies are measured by
comparing the model equation of state determined by the minimizing
parameters, $\upsilon_a$, with the GM1L equation of state used
to compute the exact $\{\tilde M_i,\tilde R_i\}$ data.  These
comparisons are made by evaluating the equation of state error
function $\Delta$ defined in Eq.~(\ref{e:Delta_EOS}) for each
$\{M_i,R_i\}$ data set.  Figure~\ref{e:Delta_Average} illustrates
$\bar\Delta$, the average value of $\Delta$ over each of the
collections, $\mathcal{A}=\{0.2,0.1, 0.01,0.001\}$, along with the
results for the exact $\mathcal{A}=0.0$ case, as functions of the
number of spectral parameters, $N_\mathrm{parms}$.  This figure shows
that increasing the number of spectral parameters does not necessarily
increase the accuracy of the equation of state determined by the
inverse stellar structure problem using noisy mass-radius data.  For
the $\mathcal{A}=\{0.2,0.1\}$ collections, the optimal number of
spectral parameters is $N_\mathrm{parms}=1$.  Observational data with
higher accuracies can benefit, however, from higher order spectral
representations.  The best accuracy for the $\mathcal{A}=0.01$
collection could perhaps be improved a little over the
$N_\mathrm{parms}=1$ case using $N_\mathrm{parms}=3$, and the accuracy
for the $\mathcal{A}=0.001$ collection could definitely be improved
over the $N_\mathrm{parms}=1$ case using $N_\mathrm{parms}=4$.

\begin{figure}[!h]
    \includegraphics[width=0.4\textwidth]
                    {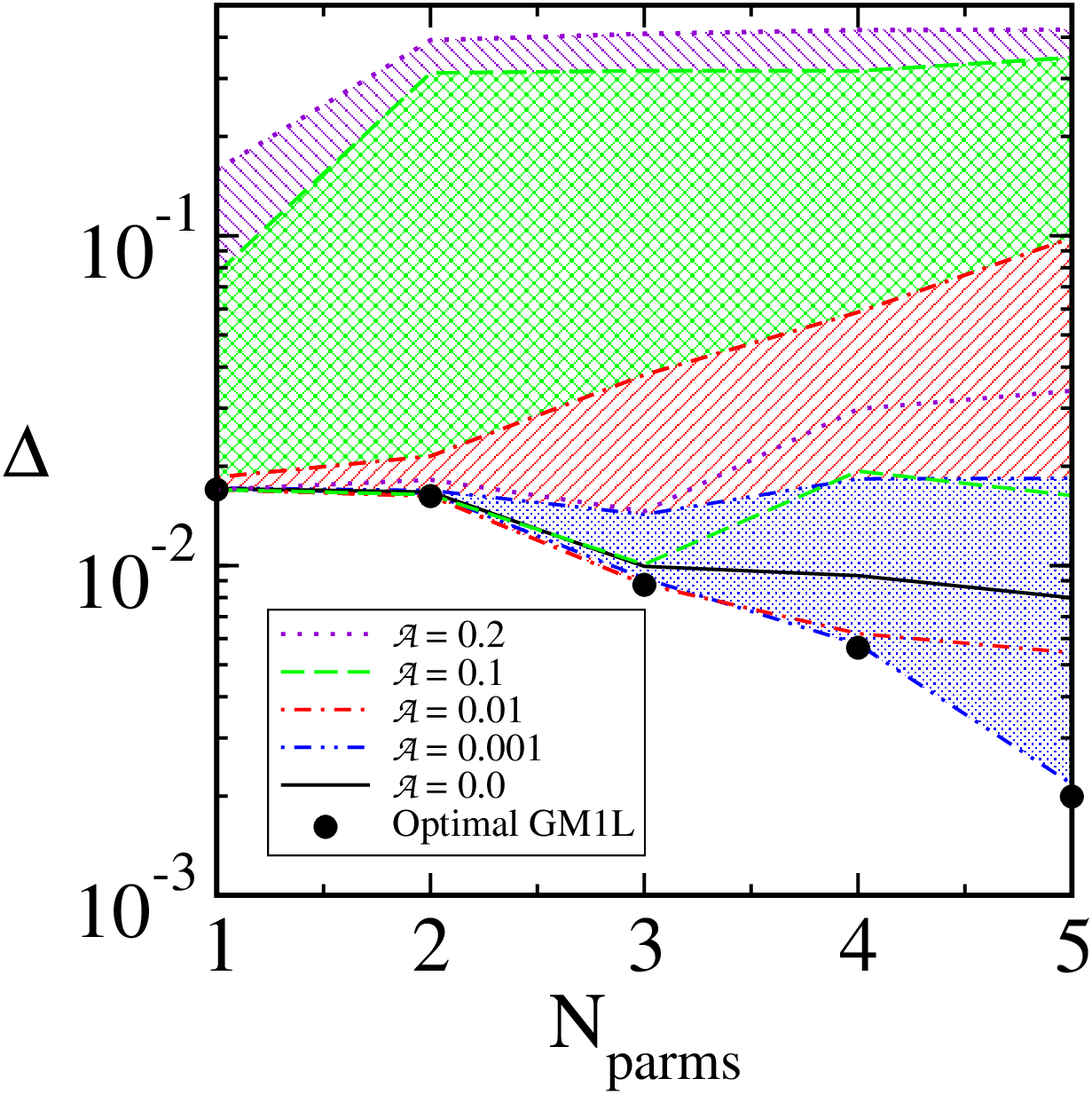}
          \caption{Ranges of the values of $\Delta$, the equation of
            state fitting error defined in Eq.~(\ref{e:Delta_EOS}),
            for the high density equation of state determined by the
            best fits for each of the perturbed $\{M_i,R_i\}$ data
            sets.  Also shown as the circular data points is the
            equation of state error $\Delta$ associated with the
            optimal spectral fit to the exact GM1L equation of state
            from which the noisy data sets were constructed for this
            study.
          \label{f:Delta_Ranges}}
\end{figure}

Figure~\ref{f:Delta_Ranges} illustrates the range of $\Delta$ values
obtained for the collections of noisy $\{M_i,R_i\}$ data with
$\mathcal{A}=\{0.2, 0.1, 0.01, 0.001\}$ as well as the values obtained
for the $\mathcal{A}=0.0$ case using the exact $\{\tilde M_i,\tilde
R_i\}$ data.  Also shown for comparison in Fig.~\ref{f:Delta_Ranges}
are the values of $\Delta$ for the optimal spectral representations of
the GM1L equation of state.  These optimal representations are
obtained by minimizing $\Delta$ defined in Eq.~(\ref{e:Delta_EOS})
with respect to variations in the equation of state parameters
$\upsilon_a$~\cite{Lindblom2024b} using all the points
$\{\epsilon_k,p_k\}$ in the table that defines the GM1L equation of
state.  These optimal values are distinct from and smaller than the
values obtained for the $\mathcal{A}=0.0$ case by minimizing $\chi$
using the exact $\{\tilde M_i,\tilde R_i\}$ data. The upper limits of
the ranges of $\Delta$ in the $\mathcal{A}=\{0.2,0.1,0.01\}$ data
collections are smallest for the $N_\mathrm{parms}=1$ spectral
representations, showing that this is the best spectral order to use
for these cases.  These upper range limits show that the equation of
state can be determined at an accuracy level commensurate with the
size of the observational error using an $N_\mathrm{parms}=1$ spectral
representation, unless the observational error is smaller than
$\mathcal{A}=0.01$.  The results in Fig.~\ref{f:Delta_Ranges} also
show that improving the accuracy with which the equation of state can
be determined by adding spectral parameters beyond
$N_\mathrm{parms}=1$ will not succeed unless the error levels in the
observational $\{M_i,R_i\}$ data are at the $\mathcal{A}=0.001$ level
or smaller.  It is also interesting to note that the minimum range of
the $\mathcal{A}=0.001$ data in Fig.~\ref{f:Delta_Ranges} extend
almost all the way to the optimal GM1L values.  Therefore some
combinations of mass-radius errors provide more accurate solutions to
the inverse structure problem than the exact mass-radius data.
Unfortunately it is not possible to know a priori what those optimal
mass-radius error combinations are for any given equation of state.
%

\section{Discussion}
\label{s:Discussion}

This study has evaluated the accuracy with which the neutron-star
equation of state can be determined by solving the inverse
relativistic stellar structure problem using noisy mass and radius
data.  Large collections of mock observational mass-radius data were
prepared by adding random errors with a range of sizes,
$\mathcal{A}=\{0.2,0.1,0.01,0.001\}$, to the exact mass-radius values
computed from a known neutron-star equation of state.  The inverse
stellar structure problem was solved using these mock data, and the
accuracy of the resulting equations of state were evaluated by
comparing them to the nuclear-theory equation of state used to
construct the mock mass-radius data.  These tests show that in most
cases the equation of state can be determined at an accuracy level
commensurate with the accuracy of the mass-radius observations.  The
exceptions to this basic result were for very high accuracy
mass-radius data, $\mathcal{A}\leq 0.001$, where the accuracy of the
equation of state could be limited by the accuracy of the equation of
state representation rather than the accuracy of the mass-radius data.

The method described in Sec.~\ref{s:InverseProblem} for solving the
relativistic inverse stellar structure problem is simpler than would
be required to analyze real observational data.  For example, the
quantity $\chi$ defined in Eq.~(\ref{e:chi_MR}) to measure how well
the computed observables $\{M_i(p_c^i,
\upsilon_a),R_i(p_c^i,\upsilon_a)\}$ agree with the observed
quantities $\{M_i,R_i\}$ treats each observed quantity equally.  The
analysis of real $\{M_i,R_i\}$ data, however, should weight each
observed quantity in $\chi$ individually, so that accurate
measurements contribute more to $\chi$, while those with less accuracy
contribute less.  The mock data constructed for this study have error
distributions of the same size for all the observables in each mock
data set.  Consequently no observable specific weighting was needed in
the definition of the $\chi$ measure for this study.  Similarly, the
errors associated with real $\{M_i,R_i\}$ observations will satisfy
complicated probably distributions determined by the particular
observational technique.  The uniform error distributions used in this
study were chosen for their simplicity.  Our expectation is that the
commensurate relationship between the size of the observational errors
and the accuracy of the inferred equation of state found here should
apply qualitatively for realistic observational error distributions.

One interesting, and perhaps somewhat counter intuitive, finding of
this study is the fact that increasing the number of parameters
included in the spectral representation does not in general increase
the accuracy of the equation of state determined by the inverse
stellar structure problem using noisy mass-radius data.  The average
equation of state errors, $\bar \Delta$, are smallest for the
$N_\mathrm{parms}=1$ representations of the noisy mass-radius data
with error amplitudes $\mathcal{A}=\{0.2,0.1\}$.  Increasing
$N_\mathrm{parms}$ in these cases produces worse approximations.  This
non-convergent behavior can occur when attempting to fit noisy data
with models having too many parameters.\footnote{Consider the simple
case of fitting an exact straight line in two-dimensions using three
noisy data points located around this line.  Fitting these data with
two parameters would produce a line whose location and slope
approximate the exact line at some level of accuracy.  Using three
parameters, however, would produce a parabola that exactly matches the
data but could be a much worse model of the exact line.  The broad
regions occupied by the large error-amplitude noisy mass-radius data
in Fig.~\ref{f:MR_DATA_RANGES} do not provide enough structure to
determine any higher-order spectral parameters beyond
$N_\mathrm{parms}=1$ for the $\mathcal{A}=\{0.2,0.1\}$ data sets.} The
exceptions to this general rule are cases with much higher accuracy
data, $\mathcal{A}=\{0.01,0.001\}$.  The $\mathcal{A}=0.01$ case is
marginal, but $\bar\Delta$ is decreased somewhat by increasing from
$N_\mathrm{parms}=1$ to $N_\mathrm{parms}=3$ in this case.  The
$\mathcal{A}=0.001$ case is more clear cut; $\bar\Delta$ is
significantly reduced by increasing from $N_\mathrm{parms}=1$ to
$N_\mathrm{parms}=3$ or $4$.

The physical neutron-star equation of state is still unknown at this
point.  Therefore in the analysis of actual mass-radius observations
it is not possible to evaluate a direct measure of the accuracy of the
inferred equation of state with the $\Delta$ measure used here.
Consequently it is more difficult in real situations to determine the
optimal number of equation of state parameters, $N_\mathrm{parms}$, to
use for the data analysis.  The results of this study, however,
suggest a possible way to do this.  Consider the $N_\mathrm{parms}$
dependence of the average mass-radius fitting errors, $\bar\chi$,
illustrated in Fig.~\ref{f:chi_Average}.  The values of $\bar\chi$ in
this study are essentially independent of $N_\mathrm{parms}$ for the
$\mathcal{A}=\{0.2,0.1,0.01\}$ collections of mock mass-radius data.
Thus the quality of the model fits to those mass-radius data are not
significantly improved beyond the $N_\mathrm{parms}=1$ fits.  This
result is in good agreement with the average accuracies, $\bar\Delta$,
of the equation of state fits shown in Fig.~\ref{e:Delta_Average}
using those mock data.  Only the $\mathcal{A}=0.001$ curve in
Fig.~\ref{f:chi_Average} shows a significant improvement in $\bar\chi$
at $N_\mathrm{parms}=3$, in good agreement again with the average
accuracy $\bar\Delta$ obtained for that case. This suggests that the
analysis of real mass-radius data should include evaluating the
mass-radius fitting errors $\chi$ for a range of values of
$N_\mathrm{parms}$ using a convergent representation of the equation
of state.  This study suggests that the optimal choice of
$N_\mathrm{parms}$ is the point where $\chi$ becomes relatively
constant for higher values of $N_\mathrm{parms}$.  Using this approach
requires the use of a parametric equation of state representation
whose accuracy can be improved by increasing the number of parameters,
e.g. those in Refs.~\cite{Lindblom2010, Lindblom2018, Lindblom2022,
  Lindblom2024b}, rather than a representation having a fixed number
of parameters like Ref.~\cite{Read:2008iy}.

Another detail that deserves a little more discussion is the
difference between the $\mathcal{A}=0.0$ results shown in
Figs.~\ref{f:chi_Average}--\ref{f:Delta_Ranges}, and the optimal GM1L
spectral fit shown in Fig.~\ref{f:Delta_Ranges}.  The
$\mathcal{A}=0.0$ results are solutions to the inverse stellar
structure problem using the exact $\{\tilde M_i,\tilde R_i\}$ data.
(The ``exact'' $\{\tilde M_i,\tilde R_i\}$ used here have fractional
errors below $10^{-7}$.)  The $\mathcal{A}=0.0$ curves in the various
figures illustrate how well the inverse structure problem could be
solved if observational errors were significantly reduced below
present levels.  The results for the optimal GM1L spectral fit shown
in Fig.~\ref{f:Delta_Ranges} represent the accuracy of the spectral
fit obtained by minimizing $\Delta(\upsilon_a)$ in
Eq.~(\ref{e:Delta_EOS}) with respect to variations in the spectral
parameters $\upsilon_a$.  The difference between the $\mathcal{A}=0.0$
and the optimal GM1L curves in Fig.~\ref{f:Delta_Ranges} shows that
there are (generally small) limitations to the accuracy of the
solutions of the inverse structure problem that go beyond the errors
in the observational data or the intrinsic accuracy of the spectral
representations.  These differences might be caused by the limited
number of data $N_\mathrm{MR}$ or the distribution of mass-radius data
used in this study.

\appendix
\section{Causal Spectral Representations}
\label{s:CausalRepresentations}
 
This Appendix summarizes the Chebyshev based causal spectral
representations of the high-density neutron-star equations of state
used in this study, developed originally in Ref.~\cite{Lindblom2024b}.
These Chebyshev based representations have been shown to provide
efficient convergent representations for a wide range of potential
neutron-star equations of state, including those with first- and
second-order phase transitions.\footnote{Spectral representations were
tested in Ref.~\cite{Lindblom2024} for model neutron-star equations of
state with a wide range of first- or second-order phase
transitions. The range of first-order transitions extended to the
discontinuity large enough to trigger an instability that terminates
the contiguous sequence of stable neutron stars.  The range of
second-order transitions extended to the upper limit set by causality.
Lower order spectral representations of equations of state with these
first- or second-order phase transitions were shown in
Ref.~\cite{Lindblom2024} to be more accurate than a non-smooth
piecewise-analytic causal representation with the same numbers of
adjustable parameters.}  The speed of sound, $v$, in a barotropic
fluid is determined by the equation of state:
$v^2=dp/d\epsilon$~\cite{Landau1959}.  These sound speeds are causal
if and only if the velocity function $\Upsilon$,
\begin{equation}
  \Upsilon=\frac{c^2-v^2}{v^2},
    \label{e:Upsilon_Def}
\end{equation}
is non-negative, $\Upsilon\geq 0$, where $c$ is the speed of light.

The velocity function $\Upsilon$ is determined by the equation of
state: $\Upsilon(p)=c^2\,d\epsilon/dp -1$.  Conversely, $\Upsilon(p)$
can be used as a generating function from which the standard equation
of state, $\epsilon=\epsilon(p)$, can be determined by quadrature.
The definition of the velocity function $\Upsilon(p)$ can be
re-written as the ordinary differential equation,
\begin{equation}
  \frac{d\epsilon(p)}{dp} =\frac{1}{c^2}+\frac{\Upsilon(p)}{c^2}.
  \label{e:Upsilon_p_Def}
\end{equation}
This equation can then be integrated to determine the equation of
state, $\epsilon=\epsilon(p)$:
\begin{equation}
  \epsilon(p)=\epsilon_\mathrm{min}+\frac{p-p_\mathrm{min}}{c^2} 
  +\frac{1}{c^2}\int_{p_\mathrm{min}}^p\Upsilon(p')dp'.
  \label{e:epsilon_p}
\end{equation}

Causal parametric representations of the neutron-star equation of
state can be constructed by expressing $\Upsilon(p,\upsilon_a)$ as a
spectral expansion based on Chebyshev polynomials developed in
Ref.~\cite{Lindblom2024b}:
\begin{equation}
  \Upsilon(p,\upsilon_a)=\Upsilon_\mathrm{min}\exp\left\{
  \sum_{a=0}^{N_\mathrm{parms}-1}\upsilon_a(1+y)T_a(y)\right\},
  \label{e:Upsilon_p_Chebyshev}
\end{equation}
where the $T_a(y)$ are Chebyshev polynomials.  The variable $y$
(defined below) is a function of the pressure having the property that
$y=-1$ when $p=p_\mathrm{min}$. The constants $p_\mathrm{min}$ and
$\Upsilon_\mathrm{min}$ are evaluated from the low-density equation of
state at the point $p=p_\mathrm{min}$ where it matches onto the high
density spectral representation determined by
Eq.~(\ref{e:Upsilon_p_Chebyshev}).  Choosing $p_\mathrm{min}$ and
$\Upsilon_\mathrm{min}$ in this way ensures that no artificial first-
or second-order phase-transition discontinuity is introduced at the
matching point.  These expansions guarantee that $\Upsilon(p)\geq 0$
for every choice of $\upsilon_a$. Therefore any equation of state
determined from one of these $\Upsilon(p,\upsilon_a)$ automatically 
satisfies the causality and thermodynamic stability conditions.

Chebyshev polynomials are defined by the recursion relation
$T_{a+1}(y)=2yT_a(y)-T_{a-1}(y)$ with $T_0(y)=1$ and $T_1(y)=y$.
Spectral expansions using Chebyshev basis functions are well behaved
on the domain $-1\leq y \leq 1$~\cite{Boyd1999}.  Therefore the
variable $y$ that appears in Eq.~(\ref{e:Upsilon_p_Chebyshev}) has
been defined as
\begin{equation}
  y = -1 + 2\log\left(\frac{p}{p_\mathrm{min}}\right)\left[
    \log\left(\frac{p_\mathrm{max}}{p_\mathrm{min}}\right)\right]^{-1},
  \label{e:zDef}
\end{equation}
to ensure that $-1\leq y\leq 1$ for pressures in the range
$p_\mathrm{min}\leq p \leq p_\mathrm{max}$.  The factor $1+y$ that
appears in Eq.~(\ref{e:Upsilon_p_Chebyshev}) ensures that
$\Upsilon(p,\upsilon_a)$ has the limit,
$\Upsilon(p_\mathrm{min},\upsilon_a)=\Upsilon_\mathrm{min}$, for every
choice of spectral parameters $\upsilon_a$.  The values of the
equation of state parameters used in this study to model the
high-density portion of the GM1L equation of state are
$p_\mathrm{min}=1.20788\times 10^{32}$ erg/cm${}^3$,
$p_\mathrm{max}=8.87671\times 10^{35}$ erg/cm${}^3$,
$\epsilon_\mathrm{min}=5.08587\times 10^{13}$ g/cm${}^3$, and
$\Upsilon_\mathrm{min}=277.532$.

\vspace{0.5cm}

\acknowledgments

We thank Fridolin Weber for providing us with the GM1L equation of
state table used in this study and L.L. thanks Steve Lewis for
numerous conversations concerning this work.  This research was
supported in part by the National Science Foundation grants 2012857
and 2407545 to the University of California at San Diego,
USA. T.Z. was supported by a Haverford College KINSC Summer Scholars
fellowship.

\vfill\break

\bibstyle{prd}
\bibliography{../References/References}

\end{document}